\documentclass[
    reprint,
    amsmath,amssymb,
    aps,
    floatfix,
]{revtex4-2}

\usepackage{graphicx}
\usepackage{xcolor}
\usepackage{hyperref}
\hypersetup{
  colorlinks=true,
  linkcolor=blue, 
  citecolor=violet,
  urlcolor=black
}

\begin{document}

\title{Disentangling epidemics and percolation}

\author{Takayuki Hiraoka}
\email{takayuki.hiraoka@aalto.fi}
\author{Jari Saramäki}
\affiliation{%
Department of Computer Science, Aalto University, Espoo 00076, Finland
}

\begin{abstract}
Non-recurrent epidemics on complex networks, such as the paradigmatic susceptible-infected-recovered (SIR) model, are often studied by mapping the model to bond percolation. However, the conditions under which this identification is justified have not been fully understood, except in specific cases. Here, we systematically determine when this mapping can be applied, starting from minimal assumptions on the epidemic model and the underlying network. Specifically, we discuss the conditions for three different forms of equivalence between a general non-recurrent epidemic model and Bernoulli bond percolation: isomorphism, equality of marginal infection probabilities, and equality of expected epidemic size. We numerically demonstrate the validity of our findings and examine how epidemiological outcomes are affected when these equivalences are not upheld. Our results establish the methodological basis and limitations of a common tool for analyzing epidemic processes on networks.
\end{abstract}

\maketitle

\section{Introduction}
One of the most notable contributions of network science is its application to modeling the spread of infectious diseases. The network-based approach provides a flexible framework for modeling epidemics at various levels of granularity; at the most granular scale, it can explicitly account for individuals and their interactions. This allows network epidemic models to readily incorporate physiological, immunological, behavioral, and social variations among hosts. 

A key theoretical development in network epidemiology was brought about by mapping epidemic processes onto percolation, a classic model in statistical physics. In standard Bernoulli bond percolation, each edge in a network is present independently with equal probability, analogous to stochastic transmission of disease from a node to its neighbor. In fact, in the first paper to formulate percolation as a mathematical model, Broadbent and Hammersley~\cite{broadbent1957percolation} cited blight spreading among trees in an orchard as a motivating example. Since then, the analogy between epidemics and percolation has been theoretically explored on lattices~\cite{welsh1977percolation, hammersley1980percolation, grassberger1983critical, kesten1987percolation}. Newman~\cite{newman2002spread} extended this link to random graphs with arbitrary degree distributions. The notion that the final outcome of a susceptible-infected-recovered (SIR) process on a network can be expressed as a connectivity problem, along with the generating-function technique for computing the epidemic threshold and outbreak sizes, has led to the widespread adoption of bond percolation as a simple epidemic model. 

Despite the apparent resemblance, however, mapping between SIR epidemics and Bernoulli bond percolation is not exact in general~\cite{mollison1977spatial, kuulasmaa1982spatial, grassberger1983critical, kesten1987percolation, kenah2007second, tome2010critical, miller2016equivalence}. For example, transmission events from an infected node to its neighbors may be correlated due to the random duration of infection: A node that happens to remain infectious longer is more likely to infect all of its neighbors. Studies have argued that SIR epidemics can only be mapped exactly to Bernoulli bond percolation if the duration of infection is degenerate. Instead, they are more accurately represented by another percolation model called \emph{epidemic percolation}, which, unlike Bernoulli bond percolation, operates on directed networks~\cite{kuulasmaa1982spatial, kesten1987percolation, kenah2007second, miller2007epidemic, trapman2007analytical}. 

However, these previous findings are based on specific epidemic processes---typically the SIR model with independent contact and recovery---and specific network structures, such as lattices or the configuration model. While these premises are mathematically convenient, they may not apply to real-world structures and dynamics. The question of how the relationship between epidemics and percolation holds under broader conditions remains open.

In this work, we explore three different types of equivalence between the final-state outcomes of a general non-recurrent epidemic model and Bernoulli bond percolation. We base our analysis on a minimal set of assumptions about the epidemic process and the underlying network. 
First, we identify the conditions under which the two models are exactly equivalent in an arbitrary single network (type-1 equivalence). 
Next, we examine conditions for a weaker form of equivalence between the two models, namely, equality of the marginal probability that each node becomes infected in an outbreak in an arbitrary single network (type-2 equivalence). We do so by comparing the message passing formalism for Bernoulli bond percolation and epidemic percolation. 
Using a similar approach, we elucidate the conditions for a third type of equivalence: equality of the expected size of a major epidemic in large configuration model networks (type-3 equivalence). 
Finally, we numerically evaluate discrepancies between the two models when these forms of equivalence do not hold.

\section{Non-recurrent epidemic model}
Many infectious diseases induce long-term or even lifelong immunity in recovered hosts, which protects them against future reinfection. The SIR model is a paradigmatic compartmental model for such non-recurrent infections. While originally formulated as a system of differential equations describing disease dynamics in a population, the SIR model, like other compartmental models, can naturally be extended to a network setting. Individuals in the population are represented by nodes in a contact network. At any given time, each node is in one of three epidemiological states: susceptible, infected, or recovered. An infected node may come into contact with its susceptible neighbors and transmit the disease, turning them into infected nodes. An infected node eventually recovers, i.e., spontaneously transitions to the recovered state, after which it becomes immune to the disease and will never become infected again nor infect others. 

\begin{figure}[bt]
    \centering
    \includegraphics[width=\linewidth]{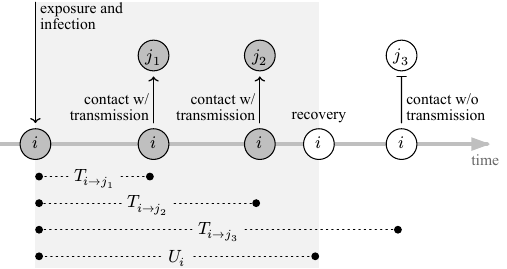}
    \caption{An illustrative example of the natural history of infection.  A susceptible node $i$ is exposed by an infected neighbor, which leads to its infection. Then, $i$ makes transmissible contacts with three susceptible neighbors: $j_1$, $j_2$, and $j_3$. The first two contacts lead to transmission, but the contact with $j_3$ does not because $i$ recovers before the contact. The shaded area represents the infected period of $i$.}
    \label{fig:natural_history}
\end{figure}

Microscopically, the dynamics of the SIR model is governed by how long it takes for an infected node to transmit the disease to others, and how quickly it recovers. A contact between two individuals is said to be \emph{transmissible} if it would result in transmission should one individual be infectious and the other be susceptible. 
Suppose that node $i$ becomes infected during an outbreak. If the next transmissible contact that $i$ makes with an adjacent node $j$ occurs at time $T_{i \to j}$ after the infection, $T_{i \to j}$ is called the infection-to-contact interval from $i$ to $j$.
We also let $U_i$ denote the time interval between the infection and recovery (i.e., the end of infectiousness) of $i$, and call it the infection-to-recovery interval. 
See Fig.~\ref{fig:natural_history} for a schematic of the model.

Node $i$ can infect adjacent node $j$ only if a transmissible contact occurs between them before node $i$ recovers, i.e., $T_{i \to j} < U_i$. We define the following indicator variable to represent whether contact between an infected node $i$ and a susceptible node $j$ would result in transmission:
\begin{equation}
    \xi_{i \to j} := 
    \begin{cases}
    1 & \text{if } T_{i \to j} < U_i,\\
    0 & \text{otherwise}.
    \end{cases}
\end{equation}

In general, the exact durations of the infection-to-contact and infection-to-recovery intervals are unknown \emph{a priori}; therefore, we treat them as random variables. 
Given the joint probability density function of the infection-to-contact interval from infected node $i$ to susceptible neighbor $j$ and the infection-to-recovery interval of $i$, the probability that $i$ transmits the disease to $j$ is expressed as
\begin{equation}
\label{eq:prob_transmission}
\begin{aligned}
    &P(\xi_{i \to j} = 1) = \int_0^\infty \int_0^{u} P(T_{i \to j} = t, U_i = u) \,d t \,d u \\
    & = \int_0^\infty P(U_i = u) \int_0^{u} P(T_{i \to j} = t \mid U_i = u) \,d t \,d u \\
    & = \int_0^\infty P(U_i = u) P(T_{i \to j} \leq u \mid U_i = u) \,d u,
\end{aligned}
\end{equation}
Hereafter, we will simplify the notation by writing the density functions of $T_{i \to j}$ and $U_i$ as $f_{U_i}$ and $f_{T_{i \to j}}$ and the cumulative distribution function of $T_{i \to j}$ as $F_{T_{i \to j}}$.

The above formulation is very general and can be used to describe a wide range of epidemic models beyond the SIR model. For instance, it can account for infections with a latent period, which is the time lag between becoming infected and becoming infectious. In the compartmental modeling approach, such infections are usually described using the susceptible-exposed-infectious-recovered (SEIR) model, with an additional exposed (E) compartment representing the latent period. However, within our formulation, there is no need to explicitly add a new compartment. One can incorporate the notion of a latent period by conditioning $T_{i \to j}$ and $U_i$ to be both greater than a certain value. Similarly, models with multiple infected compartments with different transmission rates, or even non-compartmentalized models with continuously time-varying transmission rates, can be accommodated as long as infection is non-recurrent. Note also that, at this point, we are not bound by the typical assumption that the intervals are independent. 

A straightforward way to simulate the non-recurrent epidemic model on a contact network is to use a priority queue to track infection and recovery events. The priority queue is initialized with imported infections, i.e., infection events of nodes that become infected without transmission from other nodes in the population, sorted by time of infection. These nodes are called seed nodes. At each step, the earliest scheduled infection or recovery event is extracted from the priority queue to be realized. If the extracted event is an infection event of node $i$, $i$ becomes infected if it is still susceptible. The recovery of $i$ and the infection of each susceptible neighbor $j$ are then added to the priority queue according to the generating processes of $U_i$ and $T_{i \to j}$. 

However, this sequential, event-driven algorithm is not analytically tractable and therefore offers limited insight into the model's behavior. The idea behind mapping the epidemic model onto percolation is to obtain a static counterpart of the dynamic epidemic model that yields statistically equivalent final-state outcomes and is easier to analyze. Such statistical equivalence between two models is referred to as \emph{isomorphism}. 

To invoke the mapping to percolation, we require time homogeneity as a prerequisite: the infection-to-contact and infection-to-recovery intervals must not depend on the history of the outbreak, such as the absolute time of infection and recovery and who transmitted the disease to whom. Under this assumption, a percolation model considers a ``diluted'' version of the original contact network, which we call a \emph{transmissible network}. A transmissible network consists of all the nodes in the original contact network, but includes only transmissible edges, i.e., edges through which infection is transmitted if one end node is infectious and the other is susceptible. Given a set of seed nodes and a transmissible network, the nodes that become infected in the resulting outbreak are identified as those reachable from at least one seed by traversing the network.

Establishing a link between an epidemic model and a percolation process allows us to compute various features of epidemics without having to deal directly with the dynamics, but instead by analyzing the static structure of the transmissible network.
In the following section, we will investigate the conditions under which we can leverage such a link.

\section{Isomorphism}
\label{sec:isomorphism}
Our first question is: Under what conditions are the non-recurrent epidemic model and Bernoulli bond percolation isomorphic? In Bernoulli bond percolation, each edge in an undirected network is independently retained with probability $q$ or removed with probability $1-q$. For a non-recurrent epidemic model to be isomorphic to Bernoulli bond percolation, it must satisfy the following three properties:
\begin{itemize}
\item \emph{Reciprocity}: For the epidemic model to be mapped to percolation on an \emph{undirected} network, the transmission must be reciprocal on each edge, that is, 
\[ P(\xi_{i \to j} = \xi_{j \to i}) = 1\]
for each pair of adjacent nodes $i$ and $j$.

\item \emph{Independence}: For the epidemic model to be mapped to \emph{Bernoulli} percolation, the transmissions between different ordered pairs of adjacent nodes must be independent of each other, that is, 
\[ P(\xi_{i \to j}, \xi_{k \to l}) = P(\xi_{i \to j}) P(\xi_{k \to l}) \]
for any ordered pairs of adjacent nodes $(i, j)$ and $(k, l)$.

\item \emph{Equiprobability}: For the epidemic model to be mapped to a percolation process where each edge is retained with \emph{equal probability}, the transmission probabilities between all ordered pairs of adjacent nodes must be equal, that is, 
\[ P(\xi_{i \to j} = 1) = q \]
for any ordered pair of adjacent nodes $i$ and $j$. 
\end{itemize}

In practice, the reciprocity condition is unnecessarily restrictive. Since we consider a non-recurrent infection, transmission from one node to another precludes transmission in the opposite direction. In other words, each undirected edge can be used for transmission at most once in either direction. This implies that the equality $\xi_{i \to j} = \xi_{j \to i}$ is not necessary as long as the edge is equally transmissible in both directions. Let us replace it with a more lenient condition, which is: 
\begin{itemize}
\item \emph{Symmetry}: For the epidemic model to be mapped to percolation on an \emph{undirected} network, the transmission probability must be symmetric for each edge, that is, 
\[ P(\xi_{i \to j} = 1) = P(\xi_{j \to i} = 1) \]
for each pair of adjacent nodes $i$ and $j$.
\end{itemize}

We note that equiprobability is a sufficient condition for symmetry. Therefore, in practice, only independence and equiprobability are needed. Nevertheless, it is worthwhile to make the symmetry condition explicit in order to allow for generalization to a percolation model in which edges can be retained with varying probabilities (e.g., in weighted networks).

These conditions on transmission by themselves do not fully determine the infection-to-contact and infection-to-recovery intervals. Previous studies have shown that for an SIR model to be isomorphic to bond percolation, the infection-to-recovery interval must be constant and equal for every node. However, as we will see below, this is based on a specific set of assumptions. Before getting to that, let us first examine what we can deduce about these intervals under more relaxed assumptions.

First, we assume that each infection-to-contact interval $T_{i \to j}$ is only dependent on the infection-to-recovery interval $U_i$ of the infected node $i$ and nothing else. This implies that the infection-to-contact intervals from a node to its adjacent nodes, $T_{i \to j_1}$ and $T_{i \to j_2}$, are conditionally independent given $U_i$. We do not impose any constraints on the infection-to-recovery intervals; they may be arbitrarily correlated with each other. 
Accordingly, the infection-to-contact intervals between different pairs, $T_{i \to j}$ and $T_{k \to \ell}$, are conditionally independent given $U_i$ and $U_k$. 
Under this setup, we can easily find an example that meets the three isomorphism conditions: any $T_{i \to j}$ with cumulative distribution function $F_{T_{i \to j}}$ that satisfies 
\begin{equation}
\label{eq:condition_q}
F_{T_{i \to j}}(u \mid U_i = u) = q
\end{equation}
will suffice. Importantly, the distribution of infection-to-recovery intervals does not need to be degenerate---or even independent or identically distributed---so long as it satisfies Eq.~\eqref{eq:condition_q}. 

Next, let us tighten the constraints and assume that each $T_{i \to j}$ is independent of all other variables (but not necessarily identically distributed), with the very mild condition that its cumulative distribution function $F_{T_{i \to j}}$ is strictly increasing on the support of $U_i$. This condition is equivalent to $T_{i \to j}$ having a positive probability on every interval between two possible values of $U_i$. For example, any infection-to-contact interval distribution with support $[0, \infty)$ automatically satisfies it. 
Under this assumption, the only kind of infection-to-recovery interval distribution that satisfies the isomorphism conditions is the degenerate one, that is, 
\begin{equation}
\label{eq:condition_degenerate}
f_{U_i} (u) = \delta(u - u_i^*),
\end{equation}
where $\delta(\cdot)$ denotes the Dirac delta function and $u_i^*$ is the only value that $U_i$ takes (proof in Appendix~\ref{app:proof}). In plain terms, the infection-to-recovery interval of each node must be constant. 
If $f_{U_i}$ has nonzero variance, then transmission events $\xi_{i \to j_1}, \xi_{i \to j_2}, \dots$ are not independent even though infection-to-contact intervals $T_{i \to j_1}, T_{i \to j_2}, \dots$ are, because the probability of transmission from $i$ to all of its susceptible neighbors increases with the infection-to-recovery interval $U_i$. 
Note that this does not imply that the infection-to-recovery intervals of different nodes must be equal. In fact, the constant $u_i^*$ may depend on $i$ as long as 
\begin{equation}
\label{eq:condition_q2}
    F_{T_{i \to j}}(u_i^*) = q
\end{equation}
is met for all pairs of adjacent nodes $i$ and $j$. 

Finally, in addition to the independence of $T_{i \to j}$ as in the previous case, assume that all $T_{i \to j}$ follow the same distribution irrespective of $i$ and $j$---that is, the infection-to-contact intervals are independent and identically distributed (i.i.d.). Under this rather strong i.i.d.\ assumption---and only in this case---must we require $u_i^*$ to be equal for every node $i$ in order to satisfy the isomorphism conditions. This includes the model where the contacts on each edge are described by an independent Poisson process, either with the same rate for all edges~\cite{kuulasmaa1982spatial} or with i.i.d.\ rates across edges~\cite{newman2002spread, kenah2007second}. 

In summary, we have derived conditions for the non-recurrent epidemic model to be isomorphic to Bernoulli bond percolation under a hierarchy of assumptions. Under the assumption that infection-to-contact intervals are conditionally independent, Eq.~\eqref{eq:condition_q} represents a sufficient condition. If we assume full independence of the infection-to-contact intervals, degeneracy of the infection-to-recovery interval distribution for each node is necessary [Eqs.~\eqref{eq:condition_degenerate} and \eqref{eq:condition_q2}]. Finally, if the infection-to-contact intervals are i.i.d., the infection-to-recovery interval must be constant \emph{and} equal for all nodes.

When these conditions are met, we can leverage the isomorphism to calculate final epidemic outcomes. An outbreak originating from a given set of seed nodes comprises the connected components of the transmissible network that contain at least one seed. 
In a contact network described by the configuration model, a \emph{major epidemic}---defined as an outbreak that spreads to an extensive proportion of the population---can occur only if the transmissible network has a giant component, i.e., a connected component that occupies a non-vanishing fraction of nodes in the large size limit. The epidemic threshold coincides with the onset of a giant component. The probability that an outbreak grows into a major epidemic is equal to the probability that at least one seed node is in the giant component. 

\section{Equality of node infection probabilities}
\label{sec:type2}
Our argument in the previous section shows that the isomorphism between the non-recurrent epidemic model and Bernoulli bond percolation only holds under specific conditions. These conditions are rarely met in real-world epidemics.  Even under the weakest constraints, the requirement that the infection-to-contact interval distribution should scale with the infection-to-recovery interval so as to satisfy Eq.~\eqref{eq:condition_q} is far from realistic. 

In what follows, we show that a weaker type of equivalence, under which the marginal infection probability of every node is equal in the two models, may still hold even when the models are not isomorphic. To make the distinction clear, we call this \emph{type-2 equivalence}, as opposed to \emph{type-1 equivalence}, which refers to isomorphism. To identify the conditions for type-2 equivalence, we first introduce another percolation process called \emph{epidemic percolation}, which is always isomorphic to the non-recurrent epidemic model. We then compare the message passing formalism for Bernoulli bond percolation and epidemic percolation. This clarifies when and why type-2 equivalence holds and how it can be violated.

\subsection{Epidemic percolation}
Epidemic percolation is another percolation process to which the non-recurrent epidemic model can be mapped~\cite{kuulasmaa1982spatial, kuulasmaa1984spatial, kenah2007second, kenah2007networkbased, miller2007epidemic, trapman2007analytical, karrer2010message, pellis2012reproduction, ball2016reproduction}. In contrast to Bernoulli bond percolation, which operates on an undirected network, epidemic percolation considers a directed transmissible network constructed in the following steps: (i) Generate a symmetric directed network by transforming each undirected edge in the contact network, represented by an unordered pair $\{i, j\}$, into two directed edges, represented by ordered pairs $(i, j)$ and $(j, i)$. (ii) Generate an infection-to-recovery interval $U_i$ for each node and an infection-to-contact interval $T_{i \to j}$ for each directed edge $(i, j)$ according to the given generation process. (iii) Each directed edge $(i, j)$ is retained if $T_{i \to j} < U_i$ and removed otherwise. 

By construction, epidemic percolation is always isomorphic to the non-recurrent epidemic model.
This remains true regardless of any dependencies or correlations among the infection-to-contact and infection-to-recovery intervals. Epidemic percolation provides a statistically accurate description of the disease spread as long as the generative processes of these variables are specified, whether the processes are stochastic, deterministic, independent, or correlated. 

As in the case of Bernoulli bond percolation, reachability from seed nodes determines which nodes become infected in an outbreak. However, since the transmissible network is now directed, we need to distinguish reachability from node $i$ to node $j$ and reachability in the opposite direction, from $j$ to $i$. An outbreak spreads to the out-components of the seed nodes; in other words, a node becomes infected if and only if there is at least one seed node in its in-component.

When the contact network is an instance of the configuration model, the epidemic threshold is marked by the emergence of a giant strongly connected component (GSCC), which covers a non-vanishing fraction of the transmissible network. The size of a major epidemic is bounded from below by the number of nodes in the giant out-component, i.e., the set of nodes that can be reached from the GSCC. On the other hand, the probability of a major epidemic is represented by the probability that at least one seed node is in the giant in-component---the set of nodes that can reach the GSCC.

\subsection{Message passing formalism}

The message passing formalism is a framework for efficiently calculating the value of a variable associated with each node in a network, hereafter referred to as the \emph{state} of the node, by describing the dependencies between nodes. In network science, this approach has been successfully applied to a range of problems, including percolation, community detection, and spectral graph analysis~\cite{karrer2014percolation, cantwell2019message, newman2023message}. In the context of non-recurrent infection, the variable of interest will be the probability that each node remains uninfected in an outbreak.

The message passing formalism begins by defining the \emph{neighborhood} of each node~\footnote{In graph theory, the \emph{neighborhood} of a node sometimes refers to the subgraph induced by the directly adjacent nodes, which corresponds to the special case of our neighborhood with $r=3$.}. In principle, the neighborhood of node $i$ is a subgraph consisting of nodes whose state can directly influence the state of $i$. Any influence from nodes outside the neighborhood will be exerted through nodes and edges in the neighborhood. In addition, we assume that the state of each node in the neighborhood would be independent if all edges in the neighborhood were removed from the network.

In practice, such a neighborhood may be (though not necessarily) identified by assembling every node and edge that lies on cycles of up to length $r \geq 2$, starting and ending at the focal node~\cite{cantwell2019message}. The choice of the maximum cycle length $r$ reflects one's belief about how safe it is to neglect the correlation between the states of nodes in the neighborhood that is propagated through paths outside the neighborhood, given the network structure and dynamics in question. If the network in fact does not contain any cycles longer than $r$, the message passing formalism is exact.
Note that when $r = 2$, the neighborhood is defined as the focal node, the nodes directly adjacent to it, and the edges between the focal node and each adjacent node.

\begin{figure}[bt]
    \centering
    \includegraphics[width=\linewidth]{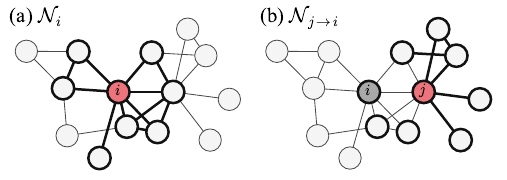}
    \caption{Schematic illustration of (a) the neighborhood of node $i$ and (b) the neighborhood of node $j$ with $i$ in the cavity state. In this example, the neighborhood of a node is defined as the subgraph consisting of nodes and edges that lie on cycles of length up to $r = 3$ starting and ending at the node. The nodes and edges in each neighborhood are indicated by bold lines.}
    \label{fig:neighborhood}
\end{figure}

Before setting out the details of the message passing formalism, let us introduce notation and concepts involved. We let $G = (V, E)$ represent the contact network. The state of node $i$ depends on the states of the other nodes in its neighborhood, which is denoted by $\mathcal{N}_i = (\mathcal{V}_i, \mathcal{E}_i)$. The state of node $j \neq i$ in $\mathcal{V}_i$ depends on the states of the other nodes in its own neighborhood, but for the sake of computing the state of node $i$, the influence of node $i$ on $j$, exerted through $\mathcal{N}_i$, should be excluded to prevent dependency ``echoing'' back and to avoid double counting the influence. Namely, we consider the modified neighborhood of $j$ where all edges in $\mathcal{N}_i$ are excluded, that is, 
$
\mathcal{N}_{j \to i} = (\mathcal{V}_j, \mathcal{E}_j \setminus \mathcal{E}_i).
$
Here, node $i$ is said to be in the \emph{cavity} state. See Fig.~\ref{fig:neighborhood} for examples of $\mathcal{N}_i$ and $\mathcal{N}_{j \to i}$. We also introduce the network $\mathcal{G}_{i} = (V, E \setminus \mathcal{E}_i)$, which is identical to $G$ except that node $i$ is in the cavity state. In other words, node $i$ is isolated from the rest of the network in $\mathcal{G}_{i}$.

By diluting the contact network $G$, we obtain a transmissible network $\tilde{G} = (V, \tilde{E})$, which is undirected for Bernoulli bond percolation and directed for epidemic percolation. In this transmissible network, the neighborhood of node $i$ maps to $\tilde{\mathcal{N}}_i = (\mathcal{V}_i, \tilde{\mathcal{E}}_i)$, where
\[ \tilde{\mathcal{E}}_i = \mathcal{E}_i \cap \tilde{E} \]
if $\tilde{G}$ is undirected and 
\[\tilde{\mathcal{E}}_i = \{(j, k) \in \tilde{E} \mid \{j, k\} \in \mathcal{E}_i\}\]
if $\tilde{G}$ is directed. In the same vein as above, we define the neighborhood of node $j$ and the full network, both with node $i$ in the cavity state, as
$
\tilde{\mathcal{N}}_{j \to i} = (\mathcal{V}_j, \tilde{\mathcal{E}}_j \setminus \tilde{\mathcal{E}}_i),
$
and
$
\tilde{\mathcal{G}}_{i} = (V, \tilde{E} \setminus \tilde{\mathcal{E}}_i).
$

With these notations, we are ready to formulate message passing equations for Bernoulli bond percolation and epidemic percolation. Let us consider the probability of each node being infected in an outbreak that spreads from a given set of seed nodes, denoted by $A$. In Bernoulli bond percolation, a node will be infected if it is in the same connected component as one of the seed nodes in the transmissible network. Node $i$ is connected to a seed node if (i) itself is a seed node, or (ii) there is a node $j$ in the neighborhood such that $j$ is connected to a seed node in the cavity network $\tilde{\mathcal{G}}_{i}$ and there is at least one path between $j$ and $i$ in $\tilde{\mathcal{N}}_i$. 

Let $\Xi_{j \to i}(\Gamma)$ be the indicator function that takes the value 1 if there is a path from node $j$ to node $i$ in network $\Gamma$ and the value 0 otherwise. For a given configuration $\Gamma$ of the neighborhood of node $i$, the probability that node $i$ is \emph{not} connected to a seed node through node $j$ in the neighborhood is given by $\mu_{j \to i}^{\Xi_{j \to i}(\Gamma)}$, where $\mu_{j \to i}$ denotes the probability that node $j$ is \emph{not} connected to any seed node in $\tilde{\mathcal{G}}_i$. Since the states of different nodes in $\mathcal{V}_i$ are assumed to be independent in $\mathcal{G}_{i}$, the probability that $i$ is \emph{not} connected to a seed node in $\tilde{G}$ is given by $\mu_i = 0$ if $i \in A$, and 
\begin{equation}
    \mu_i = \sum_\Gamma P(\tilde{\mathcal{N}}_i = \Gamma) \prod_{j \in \mathcal{V}_i \setminus \{i\}} \mu_{j \to i}^{\Xi_{j \to i}(\Gamma)}
    \label{eq:marginal_bbp_general}
\end{equation}
otherwise.
This equation represents the dependence of the state of node $i$ on the states of other nodes in the neighborhood, each of which in turn depends on the states of other nodes in its own neighborhood with $i$ in the cavity state; that is, $\mu_{j \to i} = 0$ if $j \in A$, and 
\begin{equation}
    \mu_{j \to i} = \sum_\Gamma P(\tilde{\mathcal{N}}_{j \to i} = \Gamma) \prod_{k \in \mathcal{V}_j \setminus \{j\}} \mu_{k \to j}^{\Xi_{k \to j}(\Gamma)}
    \label{eq:mp_bbp_general}
\end{equation}
otherwise.
Equation~\eqref{eq:mp_bbp_general} describes how the dependence of a node on other nodes is propagated from one neighborhood to another, hence the name ``message passing''. We can write down an equation of this form for every node $i \in V$ and for every node $j \in \mathcal{V}_i \setminus \{i\}$. The set of equations can then be numerically solved by iteration.

In epidemic percolation, the infection of a node implies that the in-component of the node contains a seed (or equivalently, the node belongs to the out-component of a seed) in the directed transmissible network. Following a similar reasoning as in the case of Bernoulli bond percolation, we can express the probability that no seed node is in the in-component of node $i$ in $\tilde{G}$ as $\mu_i = 0$ for $i \in A$ and 
\begin{equation}
    \mu_i = \sum_\Gamma P(\tilde{\mathcal{N}}_i = \Gamma) \prod_{j \in \mathcal{V}_i \setminus \{i\}} \mu_{j \to i}^{\Xi_{j \to i}(\Gamma)}
    \label{eq:marginal_epi_general}
\end{equation}
otherwise. Here, $\mu_{j \to i}$ denotes the probability that no seed node is in the in-component of node $j$ in $\tilde{\mathcal{G}}_i$. The message passing equation reads as follows: $\mu_{j \to i} = 0$ for $j \in A$ and 
\begin{equation}
    \mu_{j \to i} = \sum_\Gamma P(\tilde{\mathcal{N}}_{j \to i} = \Gamma) \prod_{k \in \mathcal{V}_j \setminus \{j\}} \mu_{k \to j}^{\Xi_{k \to j}(\Gamma)}
    \label{eq:mp_epi_general}
\end{equation}
otherwise.

\subsection{Equivalence in trees}
We immediately observe that the message passing equations for Bernoulli bond percolation and epidemic percolation are formally exactly the same---so what is the difference? Recall that, in epidemic percolation, we generate a pair of oppositely directed edges for every edge in the contact network, which may be retained or removed individually. This means that the possible configurations for epidemic percolation are as many as the square of the number of possible configurations for Bernoulli bond percolation. 

\begin{figure}[bt]
    \centering
    \includegraphics[width=\linewidth]{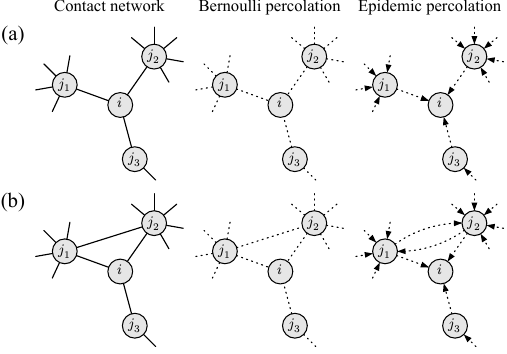}
    \caption{Schematic figure of message passing formalism for Bernoulli and epidemic percolation processes. (a) When the contact network is locally treelike, the only path from each neighbor $j$ to the focal node $i$ is through the direct edge between them. In epidemic percolation, only the incoming edges to $i$ and the incoming edges to $j$ with $i$ in the cavity state are relevant for the message passing formalism; no edges outgoing from the same node are involved. (b) When the contact network contains short cycles, such as triangles, there can be multiple paths from each neighbor $j$ to the focal node $i$. In this case, epidemic percolation involves multiple outgoing edges from $j$.}
    \label{fig:two_percolations}
\end{figure}

However, not all directed edges in epidemic percolation necessarily contribute to the message passing equations. To see this, let us consider the case where $r = 2$, meaning the neighborhood of a node consists of its adjacent nodes and incident edges. Solving the message passing equations in this case yields an exact result when the contact network is a tree.
In this case, we can simplify the message passing equations. In Bernoulli bond percolation, the edge between the focal node and each of its adjacent nodes, which is independently retained with probability $q$, is the only path between the two nodes, leading to $\mu_i = 0$ for $i \in A$ and 
\begin{equation}
    \mu_i = \prod_{j \in \mathcal{V}_i \setminus \{i\}} [1 - q + q \mu_{j \to i}] 
    \label{eq:marginal_bbp_r2}
\end{equation}
otherwise. The message passing equation is then given by $\mu_{j \to i} = 0$ for $j \in A$ and 
\begin{equation}
    \mu_{j \to i} = \prod_{k \in \mathcal{V}_j \setminus \{i, j\}} [1 - q + q \mu_{k \to j}]
    \label{eq:mp_bbp_r2}
\end{equation}
otherwise.
In epidemic percolation, the marginal probability is calculated by $\mu_i = 0$ for $i \in A$ and 
\begin{equation}
    \mu_i = \prod_{j \in \mathcal{V}_i \setminus \{i\}} [P(\xi_{j \to i} = 0) + P(\xi_{j \to i} = 1) \mu_{j \to i}].
    \label{eq:marginal_epi_r2}
\end{equation}
where $\mu_{j \to i}$ is obtained by solving the message passing equation: $\mu_{j \to i} = 0$ for $j \in A$ and 
\begin{equation}
    \mu_{j \to i} = \prod_{k \in \mathcal{V}_j \setminus \{i, j\}} [P(\xi_{k \to j} = 0) + P(\xi_{k \to j} = 1) \mu_{k \to j}]
    \label{eq:mp_epi_r2}
\end{equation}
otherwise.
Here, notice that the outgoing edges in the neighborhood $\mathcal{N}_i$ are not involved in Eq.~\eqref{eq:marginal_epi_r2}. Likewise, whether outgoing edges in the cavity neighborhood $\mathcal{N}_{j \to i}$ are retained is irrelevant in Eq.~\eqref{eq:mp_epi_r2}; see also the schematic illustration Fig.~\ref{fig:two_percolations}(a). 

For the probability $\mu_i$ to be equal in the two percolation processes, the message passing equations for epidemic percolation must reduce to those for Bernoulli bond percolation. By comparing the above four equations, we identify the conditions that the epidemic model needs to satisfy:
\begin{itemize}
\item \emph{Symmetry}: The probability that direct edges in the opposite direction are present is equal, that is, 
\[ P(\xi_{i \to j} = 1) = P(\xi_{j \to i} = 1). \]

\item \emph{Independence of incoming edges}: The presence of \emph{incoming} edges to each node is independent, that is, 
\[ P(\xi_{j_1 \to i}, \xi_{j_2 \to i}) = P(\xi_{j_1 \to i})P(\xi_{j_2 \to i}).\]
    
\item \emph{Independence of transitive edges}: The presence of \emph{transitive} edges is independent, that is, 
\[ P(\xi_{k \to j}, \xi_{j \to i}) = P(\xi_{k \to j})P(\xi_{j \to i}) \]
for $k \neq i$~\footnote{In fact, the independence of transitive edges is not a condition but rather an implicit assumption behind the message passing formalism.}.
    
\item \emph{Equiprobability}: All directed edges are present with an equal probability, that is, 
\[ P(\xi_{i \to j} = 1) = q. \] 
\end{itemize}
The schematic illustration in Fig.~\ref{fig:two_percolations}(a) depicts these conditions graphically. The conditions for type-2 equivalence (equality of node infection probabilities) are similar to those for type-1 equivalence (isomorphism), but with one notable exception: independence between outgoing edges from each node is not required. As the presence of outgoing edges is irrelevant when computing $\mu_i$ in Eqs.~\eqref{eq:marginal_epi_r2} and \eqref{eq:mp_epi_r2}, the conditions are more relaxed. 

How do these conditions translate into the constraints for contact and recovery processes? Let us address this question in the same way as we did for isomorphism---by varying the model assumption step by step. 
The weakest assumption we considered for isomorphism was that the infection-to-contact interval $T_{i \to j}$ is only dependent on the infection-to-recovery interval $U_i$ of the infected node $i$. Under this assumption, $T_{i \to j}$ following any distribution that satisfies Eq.~\eqref{eq:condition_q} achieves type-2 equivalence as well. 
When we assume that $T_{i \to j}$ is fully independent, the conditions for type-2 equivalence are met if and only if
\begin{equation}
    \int_0^\infty f_{U_i} (u) F_{T_{i \to j}}(u) \,d u = q.
    \label{eq:condition_q3}
\end{equation}
This means that the distributions of the infection-to-contact and infection-to-recovery intervals should scale with each other so that the transmission probability remains constant. Finally, Eq.~\eqref{eq:condition_q3} also serves as the necessary and sufficient condition for type-2 equivalence under the assumption that $\{T_{i \to j}\}$ are i.i.d. Compared to Eqs.~\eqref{eq:condition_degenerate} and \eqref{eq:condition_q2}, this is a more generous condition as it can accommodate an infection-to-recovery interval distribution with non-zero variance.

In sum, when the contact network is a tree, and under the independence or i.i.d.\ assumption, Bernoulli bond percolation can accurately predict the probability of each node becoming infected, even if it is not isomorphic to the epidemic model. As a corollary, the expected size of an outbreak starting from the given set of seeds, computed as the sum of node infection probabilities, is also accurate. In the next subsection, we see if this conclusion can be extended to networks with cycles.

\subsection{Equivalence in networks with cycles}
To examine the conditions for type-2 equivalence in contact networks with cycles, it is useful to distinguish between networks that contain short cycles and those that do not. By ``short'' cycles, we mean cycles short enough to be contained within the neighborhood of a node. A neighborhood can be arbitrarily large; however, for the sake of our argument about networks with short cycles, it is sufficient to consider a neighborhood that includes an edge between adjacent nodes, as shown in Fig.~\ref{fig:two_percolations}(b). Such a neighborhood is identified by $r = 3$.

In the neighborhood of node $i$, there may be more than one possible path from an adjacent node $j$ to the focal node $i$: in addition to the direct path from $j$ to $i$, there may be one or more two-step paths that pass through other adjacent nodes. These paths necessarily involve multiple outgoing edges from $j$. 
Consequently, for type-2 equivalence to hold, we must additionally require: 
\begin{itemize}
\item \emph{Independence of outgoing edges}: The presence of \emph{outgoing} edges from each node is independent, that is, 
\[ P(\xi_{i \to j_1}, \xi_{i \to j_2}) = P(\xi_{i \to j_1})P(\xi_{i \to j_2}). \]
\end{itemize}

This additional condition makes the requirements for type-1 and type-2 equivalences identical. In other words, when the contact network contains short cycles, the Bernoulli bond percolation model accurately predicts node infection probabilities only if the isomorphism holds. This conclusion generalizes to networks with cycles of finite lengths greater than three.

What if the network does not contain short cycles and is therefore locally treelike, but contains long cycles? Large Erd\H{o}s--R\'{e}nyi (ER) graphs and configuration model networks are well-known examples of such networks. It is commonly presumed that tree-based theories are asymptotically accurate in locally treelike networks. However, we show below that this presumption does not hold for the \emph{seeded} message-passing equations of the two percolation models: even when the network is locally treelike, the solution obtained with a boundary condition does not by itself give the probability of infection from the seeds.

\begin{figure}[bt]
    \centering
    \includegraphics[width=\linewidth]{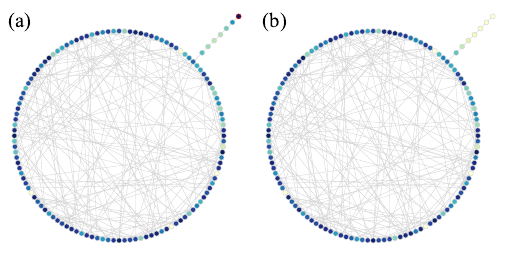}
    \caption{The solution of the message passing equations for Bernoulli bond percolation, Eq.~\eqref{eq:mp_bbp_r2}. The equations are solved (a) with a seed (enclosed in red), and (b) without a seed. Node colors represent node marginals, with lighter colors indicating larger marginal values. The network is a combination of an Erd\H{o}s--R\'{e}nyi graph with 120 nodes and a mean degree of 2.5 (shown in a circle) and a path graph with seven nodes (in the top right corner). The edge retention probability $q = 0.6$.}
    \label{fig:message_passing_seeded}
\end{figure}

Let us look at an illustrative example. Consider a network in which a path graph is connected to a large ER graph, as depicted in Fig.~\ref{fig:message_passing_seeded}. Clearly, this network is locally treelike. Let us suppose that the node at the tip of the path subgraph is the only seed node and solve Eq.~\eqref{eq:mp_bbp_r2} with the corresponding boundary condition. Figure~\ref{fig:message_passing_seeded}(a) shows that the node marginal $\mu_i$ obtained as the solution of Eq.~\eqref{eq:marginal_bbp_r2} exhibits an unexpected behavior: The values of $\mu_i$ for some nodes in the ER subgraph are smaller than those for nodes in the middle of the path subgraph. This is inconsistent with the interpretation that $\mu_i$ denotes the probability that node $i$ is not reachable from the seed in a transmissible network---since infection must be transmitted through all nodes in the path subgraph to reach the ER subgraph, it is impossible that a node in the ER subgraph becomes infected with higher probability than any node in the path subgraph.

The reason for the inconsistency is that the message passing equations, when solved with a boundary condition, capture not only reachability from the seeds but also reachability from other things. This ``other thing'' has long been thought to be the giant component---an extensive component whose size scales with the network---although recently, we argued that it is more naturally interpreted as reachability from cycles~\cite{hiraoka2026message}, where we showed that the nontrivial fixed point of the message-passing equations tracks reachability from cycles rather than from the giant component. In what follows, we adopt the latter interpretation, though this choice does not affect our conclusions. Either way, the presence of cycles, however long they may be, prevents us from giving $\mu_i$ the same straightforward interpretation it has in the case of trees. Intuitively, we can think of this as artificially self-sustained disease transmission without any imported infection of seeds. 

To correct for this artefactual behavior, we may consider solving the message passing equations \emph{without} imposing any boundary conditions---that is, without specifying seeds and hence without fixing any message to zero. Since this solution purely captures reachability from cycles, it represents the contribution that is inevitably mixed into the solution obtained \emph{with} boundary conditions at the seeds. We can therefore use this boundary-free solution to remove the unwanted ``background'' from the boundary-conditioned solution. 

For clarity, let us reintroduce the notation: We represent node marginals obtained by solving message passing equations with a set of seeds $A$ by $\mu_i(A)$, and node marginals obtained by solving them without boundary conditions by $\mu_i(\varnothing)$. We also denote the event that node $i$ is \emph{not} reachable from any seed in the transmissible network as $\mathcal{S}_i$ and the event that it is \emph{not} reachable from any cycle as $\mathcal{C}_i$. According to our argument, we have 
\begin{align}
    \mu_i(A) &= P(\mathcal{S}_i \cap \mathcal{C}_i), \\
    \mu_i(\varnothing) &= P(\mathcal{C}_i).
\end{align}
What we are truly interested in is the probability that $i$ is not infected in an outbreak, i.e., $P(\mathcal{S}_i)$, which can be calculated using the inclusion-exclusion principle as
\begin{equation}
    \begin{aligned}
        P(\mathcal{S}_i) 
        &= P(\mathcal{S}_i \cup \mathcal{C}_i) - P(\mathcal{C}_i) + P(\mathcal{S}_i \cap \mathcal{C}_i)\\
        &= 1 - P(\neg \mathcal{S}_i \cap \neg \mathcal{C}_i) - \mu_i(\varnothing) + \mu_i(A),
    \end{aligned}
    \label{eq:true_non_inf_prob}
\end{equation}
where $\neg$ denotes the complementary event. Equation~\eqref{eq:true_non_inf_prob} implies that computing the true probability of node $i$ not being infected requires calculating the probability that $i$ is reachable both from a seed ($\neg \mathcal{S}_i$) and from a cycle ($\neg \mathcal{C}_i$). 

In Bernoulli bond percolation, $\neg \mathcal{S}_i \cap \neg \mathcal{C}_i$ simply means that $i$ belongs to a multicyclic component that contains a seed. In epidemic percolation, however, this event can arise through several different scenarios. As a simplified example, let us consider an outbreak initiated by a single seed, denoted by $a$, on a contact network where there is only one multicyclic strongly connected component. If a multicyclic strongly connected component remains in the transmissible network, we represent it by $c$. Writing $j \leadsto k$ to represent that $k$ is reachable from $j$ without passing through $i$,
and noting that $\neg \mathcal{S}_i \cap \neg \mathcal{C}_i = \{a \leadsto i\} \cap \{c \leadsto i\}$, we can partition this event according to the mutual reachability of $a$ and $c$ into the following three cases. (i) $a \leadsto c \leadsto i$: the seed can reach the cycle and the cycle can reach node $i$, that is, $i$ can be reached from the seed \emph{through} the cycle. (ii) $c \leadsto a \leadsto i$ with $a \not\leadsto c$: node $i$ can be reached from the seed, which lies strictly downstream of the cycle. The condition $a \not\leadsto c$ excludes the configurations already counted in case (i). (iii) $a \leadsto i$ and $c \leadsto i$ with $a \not\leadsto c$ and $c \not\leadsto a$: the seed and the cycle are mutually unreachable and reach node $i$ through separate paths.
Here, our focus is not on actually computing the probability of each of these events, but rather on recognizing that the computation would require an assessment of whether the seed $a$ can reach the cyclic component $c$. The probability that $a$ cannot reach $c$ can be obtained by solving the following message passing equations without boundary conditions:
\begin{equation}
    \mu_a = \prod_{j \in \mathcal{V}_a \setminus \{a\}} [P(\xi_{a \to j} = 0) + P(\xi_{a \to j} = 1) \mu_{j \to a}],
    \label{eq:marginal_epi_r2_reverse}
\end{equation}
with 
\begin{equation}
    \mu_{j \to a} = \prod_{k \in \mathcal{V}_j \setminus \{a, j\}} [P(\xi_{j \to k} = 0) + P(\xi_{j \to k} = 1) \mu_{k \to j}].
    \label{eq:mp_epi_r2_reverse}
\end{equation}
Note that the subscripts of $\xi$ in these equations run in the opposite direction to those in Eqs.~\eqref{eq:marginal_epi_r2} and \eqref{eq:mp_epi_r2}. Let us observe that if these message passing equations have a nontrivial solution, they involve multiple outgoing transmissions from the same node (node $a$ or the closest node to $a$ with a degree greater than two). Plugging this observation back into Eq.~\eqref{eq:true_non_inf_prob}, we can conclude that the independence of outgoing transmissions is required for node infection probabilities to be equal in Bernoulli bond percolation and epidemic percolation on locally treelike networks with long cycles, too. In other words, the conditions for type-2 equivalence again fall back to those for type 1.

\section{Equality of expected epidemic size}
\label{sec:type3}
Type-2 equivalence concerns a node-by-node comparison of the outcomes of the two percolation processes. There, the focus is on whether Bernoulli bond percolation can accurately predict the probability that each individual will become infected. While such microscopic knowledge is valuable, we may also be interested in macroscopic (population-wide) epidemic statistics. Of particular interest is the expected final size of a major epidemic in large configuration model networks. We refer to the equivalence between the two percolation processes with respect to this quantity as \emph{type-3 equivalence}. Previous studies have demonstrated that type-3 equivalence can be achieved under the assumption of i.i.d.\ infection-to-contact intervals~\cite{newman2002spread, kenah2007second}, but this merely presents a sufficient condition. Following an approach similar to that in previous sections, we now unpack the conditions for type-3 equivalence from a more general perspective.

In the epidemic percolation framework, the size of a major epidemic is asymptotically equal to the size of the giant out-component of the transmissible network. Although there may be nodes outside the giant out-component that are reachable from one of the seed nodes, the number of such nodes becomes negligible in the large-size limit, provided that the number of seeds is finite. 

The expected size of the giant out-component can be obtained by summing the probability that each node is reachable from the GSCC. If we adopt the conjecture that reachability from the GSCC in a large configuration model network can be identified with reachability from multiple cycles~\footnote{Currently, this remains a conjecture. For the directed Erd\H{o}s--R\'{e}nyi graph, it has been shown that the GSCC is the only strongly connected component with more than one cycle in the supercritical regime~\protect\cite{luczak2009critical, dovgal2024birth}. For the directed configuration model, Coulson~\protect\cite{coulson2021strong} showed that all strongly connected components are either cycles or isolated nodes in the barely subcritical regime. However, to our knowledge, the relationship between the GSCC and the number of cycles in the supercritical regime has not yet been established, nor has it been studied in the context of epidemic percolation. Nevertheless, we can reasonably expect that this holds true based on analogy with the undirected case.}, this probability can be computed by solving Eq.~\eqref{eq:mp_epi_r2} \emph{without} a boundary condition. Similarly, the expected size of the giant component in Bernoulli bond percolation is equal to the sum of the probabilities that each node is reachable from the giant component. Since the giant component is the only component that contains multiple cycles~\cite{molloy1995critical}, we can calculate these probabilities by solving Eq.~\eqref{eq:mp_bbp_r2} without a boundary condition. Following the same line of reasoning as in the previous section, the conditions for these two sizes to coincide in expectation are symmetry, independence of incoming edges, independence of transitive edges, and equiprobability. Independence of outgoing edges is not necessary. 

Taking these two steps together leads us to conclude that the expected size of the giant component in Bernoulli bond percolation is asymptotically equal to the expected size of a major epidemic, provided that the isomorphism conditions minus the independence of outgoing transmissions are met. Therefore, when each infection-to-contact interval is independent or i.i.d., type-3 equivalence is achieved under the condition of Eq.~\eqref{eq:condition_q3}.

The conditions on the epidemic model for type-3 equivalence are exactly the same as those for type-2 equivalence in trees. Nevertheless, it is important to clarify the distinction between the two. First, these two types of equivalence arise in different classes of contact networks: In the absence of independence between outgoing transmissions, node infection probabilities are only equal in trees, while the coincidence between the expected sizes of the giant (out-)component is predicated on the contact network following the configuration model. Second, node infection probabilities are always computed against a given set of seed nodes, whereas the expected size of the giant component is agnostic to the placement of seeds and only concerns the structure of the transmission network. 

\section{Effects of equivalence breaking}
\label{sec:nonequiv}
We have discussed three different types of equivalence between the non-recurrent epidemic model and Bernoulli bond percolation. The conditions for these equivalences to hold are so specific that one could argue they are rarely met in practical settings. But how relevant is this in practice? To what extent do violations of these conditions quantitatively alter epidemiological outcomes, and how confidently can we use the Bernoulli bond percolation model as a sufficiently good approximation?

Although answering these questions in general terms is difficult, we can consider simple models of violating equivalence conditions to assess their impact. Specifically, we consider the following four cases: (a) correlations between incoming transmissions, (b) correlations between transitive transmissions, (c) correlations between outgoing transmissions, and (d) triangles in the contact network.

\subsection{Models}
Our models for the first three cases use the gamma distribution to induce correlations. The gamma distribution is defined by the probability density function
\begin{equation}
f(x; \kappa, \theta) = \frac{(x / \theta)^{\kappa-1} e^{- x / \theta}}{\Gamma(\kappa)\theta},
\end{equation}
parameterized by shape parameter $\kappa$ and scale parameter $\theta$. The shape parameter $\kappa$ controls the heterogeneity of the distribution: a smaller $\kappa$ means a larger variance. 

\paragraph{Correlations between outgoing transmissions.} 
To model correlated outgoing transmissions, we assume the following:
(i) All intervals $T_{i \to j}$ and $U_i$ are independent.
(ii) The infection-to-contact intervals $T_{i \to j}$ are drawn from an identical exponential distribution with rate $\beta$.
(iii) The infection-to-recovery intervals $U_i$ are drawn from an identical gamma distribution, parameterized by shape $\kappa$ and scaled so that the marginal probability of transmission along a random edge is equal to $q$. 

From the above conditions, we have
\begin{equation}
\begin{aligned}
    P(\xi_{i \to j} = 1) 
    &= P(T_{i \to j} < U_i) \\
    &= \int_0^\infty f_{U_i}(u) F_{T_{i \to j}}(u) \,du \\
    &= \frac{1}{\Gamma(\kappa) \theta^\kappa} \int_0^\infty u^{\kappa - 1} e^{-u / \theta} (1 - e^{-\beta u})\, du \\
    &= 1 - \left( \frac{1}{\beta \theta + 1}\right)^\kappa = q.
\end{aligned}
\end{equation}
Solving for $\theta$ yields
\begin{equation}
    \theta = \frac{(1 - q)^{-1 / \kappa} - 1}{\beta}.
\end{equation}

As $\kappa \to \infty$, the mean of the gamma distribution tends to
\begin{equation}
\begin{aligned}
    \mathbb{E}[U] = \kappa \theta = \frac{\kappa}{\beta} (e^{z/\kappa} - 1) \to \frac{z}{\beta},
\end{aligned}
\end{equation}
where we let $z = - \log(1 - q)$. Similarly, the variance tends to 
\begin{equation}
\begin{aligned}
    \mathrm{Var}[U] = \kappa \theta^2 = \frac{\kappa}{\beta^2} (e^{z/\kappa} - 1)^2 \to 0.
\end{aligned}
\end{equation}
Therefore, the distribution of infection-to-recovery intervals is degenerate and concentrated at $- \log(1 - q)/ \beta$ in the limit of $\kappa \to \infty$. Under the model constraints (i) and (ii), incoming and transitive transmissions are always independent. However, outgoing transmissions are independent if and only if infection-to-recovery intervals are constant, and decreasing $\kappa$ monotonically increases the correlations between them.

\paragraph{Correlations between incoming transmissions.}
In contrast to the correlations between outgoing transmissions, which naturally arise from independent contact and recovery processes, the correlations between incoming or transitive transmissions need to be introduced explicitly. To simulate correlated incoming transmissions, we use the following model:
(i) Each node $i$ is assigned a latent variable $w_i$, drawn independently from an identical gamma distribution with shape $\kappa$ and scale $1 / \kappa$.
(ii) Each $T_{i \to j}$ is obtained by first sampling an auxiliary infection-to-contact interval $\tilde{T}_{i \to j}$ from an exponential distribution with rate $w_j$, then transforming it so that the marginal distribution is exponential with rate $\beta$.
(iii) All $U_i$ are equal to a constant $- \log(1 - q) / \beta$, chosen so that the probability of transmission along a random edge is equal to $q$.

In this model, the parameter $\kappa$ controls the strength of correlations. The gamma distribution of $w_i$ has a mean of $1$ and a variance equal to $1 / \kappa$. As $\kappa \to \infty$, it converges to a degenerate distribution centered at 1, in which case independent sampling of $T_{i \to j}$ is recovered. On the other hand, smaller values of $\kappa$ increase the variance of the latent node variable, which induces stronger correlations between $T_{i \to j}$ with the same $j$ and, consequently, between transmissions sharing the same target node.

In step (ii), we first draw $\tilde{T}_{i \to j}$ from an exponential distribution with the gamma-distributed rate $w_j$. This variable is, in general, not exponentially distributed upon marginalization over $w_j$. Rather, it yields a Lomax distribution with cumulative distribution function (CDF)
\begin{equation}
    F_{\tilde{T}_{i \to j}}(t) = 1 - (1 + t / \kappa)^{-\kappa}.
\end{equation}
To ensure a fair comparison across values of $\kappa$, we remarginalize $\tilde{T}_{i \to j}$ to an exponential distribution using the probability integral transform. Namely, we apply its own CDF to obtain uniform variates,
\begin{equation}
    Y_{i \to j} = 1 - (1 + \tilde{T}_{i \to j} / \kappa)^{-\kappa},
\end{equation}
then map it to the target exponential distribution via its inverse CDF,
\begin{equation}
    T_{i \to j} = -\frac{1}{\beta} \log(1 - Y_{i \to j}) = \frac{\kappa}{\beta} \log(1 + \tilde{T}_{i \to j} / \kappa).
\end{equation}
By construction, $T_{i \to j}$ is exponentially distributed with rate $\beta$, while the dependence structure of $\tilde{T}_{i \to j}$ is preserved because the transformation is monotone. 

\paragraph{Correlations between transitive transmissions.}
We model correlations between transitive transmissions using a similar approach to those between incoming transmissions: 
(i) Each node $i$ is assigned a latent variable $w_i$, drawn independently from an identical gamma distribution with shape $\kappa$ and scale $1 / \kappa$.
(ii) Each $T_{i \to j}$ is obtained by first sampling $\tilde{T}_{i \to j}$ from an exponential distribution with rate $w_i + w_j$, then transforming it so that the marginal distribution is exponential with rate $\beta$.
(iii) All $U_i$ are equal to a constant $- \log(1 - q) / \beta$, chosen so that the probability of transmission along a random edge is equal to $q$.

The only difference from the model of correlated incoming transmissions is the rate of the exponential distribution from which $\tilde{T}_{i \to j}$ is sampled. The common dependence of $T_{i \to j}$ and $T_{j \to k}$ on $w_j$ induces correlation between them. Here, we exploit the fact that the sum of independent gamma-distributed random variables with the same scale value is also gamma-distributed. Namely, $w_i +w_j$ follows a gamma distribution with shape $2\kappa$ and scale $1 / \kappa$. In the same vein, we generate marginally exponential $T_{i \to j}$ through the following transformation:
\begin{equation}
    T_{i \to j} = \frac{2 \kappa}{\beta} \log(1 + \tilde{T}_{i \to j} / \kappa).
\end{equation}

In the above three models, the contact network is assumed to be an instance of the Erd\H{o}s--R\'{e}nyi random graph with $N$ nodes and mean degree $m$.

\paragraph{Networks with triangles.}
To examine the effects of short cycles, we use the random graph model proposed by Newman~\cite{newman2009random} and Miller~\cite{miller2009percolation}. This model takes as inputs the number of dyadic edges $n^\parallel_i$ and the number of triangles $n^\triangle_i$ incident to each node $i$. Viewing them as the numbers of half-edges and third-triangles that emanate from each node, the model randomly joins pairs of half-edges to form edges and triples of third-triangles to form triangles. The degree of node $i$ is given by $n_i= n^\parallel_i + 2 n^\triangle_i$. 

Our aim is to generate sequences of $n^\parallel_i$ and $n^\triangle_i$ pairs that result in variable triangle abundance while maintaining the same degree distribution for fair comparison. To this end, we first draw node degree $n_i$ independently from a Poisson distribution with a specified mean $m$. To make each of the $n_i$ incident edges constitute a triangle with probability $\rho$ and a dyadic edge otherwise, we draw the number of triangles that node $i$ is part of from a binomial distribution: 
\begin{equation}
n^\triangle_i \sim \mathrm{Binomial}(\lfloor n_i / 2 \rfloor, \tilde{\rho}),
\end{equation}
where $\lfloor \cdot \rfloor$ is a floor function, and 
\begin{equation}
\tilde{\rho} = \frac{\rho}{1 - (1 - e^{-2 m}) / 2m}.
\end{equation}
The factor $(1 - e^{-2 m}) / 2m$ accounts for the rounding of the floor function. Due to this correction, the value of $\rho$ is upper bounded by $1 - (1 - e^{-2 m}) / 2m$. Finally, the remaining $n^\parallel_i = n_i - 2 n^\triangle_i$ edges are allotted for making dyadic edges.
We require that $\sum_i n_i$ and $\sum_i n^\triangle_i$ be a multiple of two and three, respectively, so that no half-edges or third-triangles are left over.

As long as the three conditions of isomorphism are met, the structure of the contact network does not affect node infection probabilities. In other words, the presence of triangles comes into play only when one or more of these conditions is violated. For this reason, we assume that the epidemic dynamics follow the correlated outgoing transmission model, and use $\kappa = 0.2$ for simulations---a regime of strong outgoing correlation in which the equivalence is already considerably violated, so that the amplifying effect of triangles is discernible.

\subsection{Results}
Through numerical simulations of these four models, we test the impact of equivalence breaking on node infection probabilities. For the first three models, we use the same single instance of the Erd\H{o}s--R\'{e}nyi random graph with $N=10000$ and $m=3.5$ throughout, with a single fixed seed node. Because our aim is to isolate the effect of the epidemic dynamics from that of the network structure, we hold the contact network and the seed fixed across the two percolation processes, so that any discrepancy in node infection probabilities reflects the dynamics rather than variation in the substrate. We confirmed that the results are representative across network instances and seed choices. For the last model, 50 networks of $N=5000$ nodes with $m=3.5$ are used for each value of $\rho$, and a single seed node is selected for each network instance. In all cases, the (marginal) edge retention probability is $q = 0.4$, and the rate of (marginal) infection-to-contact interval distribution is $\beta = 1$. 
Empirical probabilities are computed from $10000$ independent Monte Carlo realizations of each percolation process. 

\begin{figure}[bt]
    \centering
    \includegraphics[width=\linewidth]{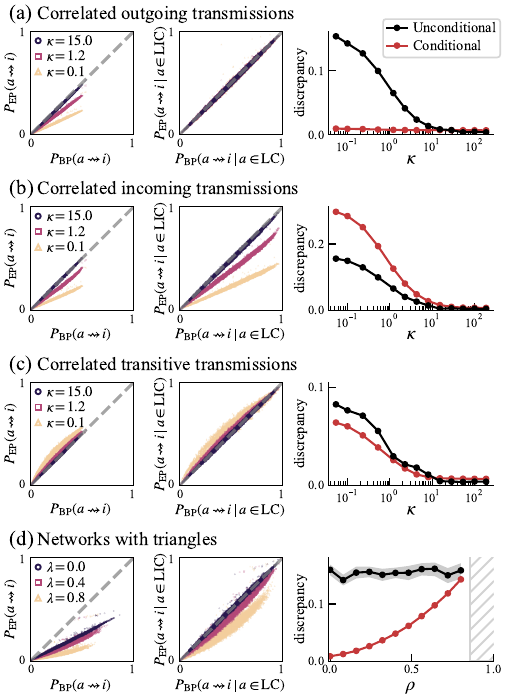}
    \caption{
        Comparison of node infection probabilities under Bernoulli bond percolation (BP) and epidemic percolation (EP). 
        Each row represents a model of equivalence breaking between the two percolation processes. 
        In each row, the left scatter plot compares the empirical probability that each node is reachable from the seed node under BP (horizontal axis) and EP (vertical axis). The dashed diagonal marks perfect agreement. The center scatter plot shows the same comparison conditioned on the seed belonging to the largest component (LC) of the transmissible network for BP and the largest in-component (LIC) for EP. 
        In (d), the two scatter plots show a random sample of 10,000 nodes (drawn from 250,000 nodes across 50 networks) to reduce overplotting.
        The right plot shows the discrepancy between the two models, measured by the mean absolute difference in unconditional (black) and conditional (red) infection probabilities, as a function of the control parameter.
        In (d), the shaded bands give the 95\% confidence interval of the discrepancy across network instances, and the hatched area denotes the inaccessible values of $\rho$. The confidence interval for the conditional curve is omitted as it is too narrow to be visible.
    }
    \label{fig:compare_bernoulli_epidemic}
\end{figure}

Figure~\ref{fig:compare_bernoulli_epidemic} summarizes the comparison between the probabilities of each node being infected (i.e., reachable from the seed node) in Bernoulli bond percolation and epidemic percolation. The discrepancy is quantified by the network-wide mean of the absolute difference of the probabilities, $\sum_i |P_\mathrm{BP}(a \leadsto i) - P_\mathrm{EP}(a \leadsto i)| / N$. 

In the correlated outgoing transmission model, shown in Fig.~\ref{fig:compare_bernoulli_epidemic}(a), the two percolation processes result in nearly identical node infection probabilities when $\kappa$ is large. Here, the conditions for equality of infection probabilities are not strictly met because the contact network is locally treelike yet contains long cycles. However, the correlation between outgoing transmissions is so weak that the presence of long cycles does not cause a discrepancy between the two percolation processes in terms of reachability from the seed node to the giant (in-)component. In other words, the two processes are almost isomorphic. 
As the value of $\kappa$ decreases and the correlation becomes stronger, the infection probabilities under epidemic percolation become increasingly smaller than under Bernoulli bond percolation. Nevertheless, the infection probabilities conditioned on the seed node being in the giant component of the transmissible network remain equal across different values of $\kappa$. In other words, each node has the same probability of being infected in the two percolation processes, provided there is a path from the seed node to the giant component. This clearly implies that the discrepancy in the raw probabilities is driven by the difference in reachability from the seed node to the giant component. 

Figure~\ref{fig:compare_bernoulli_epidemic}(b) shows that a similar behavior can be observed in the case of correlated incoming transmissions. When $\kappa$ is large and the correlation is weak, the two percolation models predict equal node infection probabilities. However, with a higher level of correlation, nodes are less likely to be reachable from the seed node in epidemic percolation than in Bernoulli bond percolation. In this case, however, we also observe a deviation in the conditional probabilities at small $\kappa$. Unlike the case of correlated outgoing transmissions, the discrepancy between the two percolation processes stems from the difference in reachability from the giant component to each node $i$. 

Correlation in transitive transmissions also creates a gap in infection probabilities between the two percolation processes [Fig.~\ref{fig:compare_bernoulli_epidemic}(c)]. Interestingly, the direction of the discrepancy is opposite to that of correlated outgoing and incoming transmissions: Correlated transitive transmissions slightly increase node infection probabilities compared to Bernoulli bond percolation. Note that because our model is based on shared latent node variables, it necessarily induces weak correlations in outgoing and incoming transmissions as well. Considering the fact that these correlations lower infection probabilities, we can conclude that the pure effect of correlation in transitive transmissions is to increase infection probabilities. 

The presence of triangles in the contact network does not itself break the equivalence conditions. However, when the conditions are already violated (e.g., by the correlation in outgoing transmissions in our model), triangles amplify the disparity between the two percolation processes, as shown in Fig.~\ref{fig:compare_bernoulli_epidemic}(d). In fact, the discrepancy, as measured by the mean absolute difference in infection probabilities, does not significantly increase with the proportion of triangles despite the visible widening of the gap in the scatter plot. This is because we are making comparisons across networks with different structures. All else being equal, large outbreaks are less likely in networks containing more triangles. Therefore, the overall probability that nodes become infected (i.e., are reachable from the seed) decreases. Due to this difference in baseline, the wider gap for nodes with a higher probability of infection is offset by the small differences for nodes with a lower probability of infection in both percolation processes. When we condition on the seed node being in the giant (in-)component, the discrepancy shows a clear increasing trend with $\rho$, which corroborates our explanation.

\begin{figure}[bt]
    \centering
    \includegraphics[width=\linewidth]{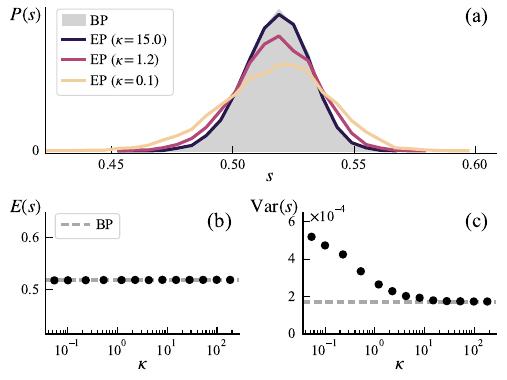}
    \caption{The statistics of the largest (out-)component size $s$ in Bernoulli bond percolation (BP) and epidemic percolation (EP) with correlated outgoing transmissions: (a) the size distribution $P(s)$, (b) the mean size $E(s)$, and (c) the variance $\mathrm{Var}(s)$.}
    \label{fig:loutc_sizes}
\end{figure}

The correlated outgoing transmission model also allows us to verify type-3 equivalence, i.e., equality of the expected major epidemic size in configuration model networks. Figure~\ref{fig:loutc_sizes} shows the comparison of the statistics of giant component sizes in Bernoulli bond percolation and giant out-component sizes in epidemic percolation. At a high value of $\kappa$, the size distributions in epidemic percolation and Bernoulli bond percolation are in agreement. Consequently, the mean major epidemic size coincides with the size of the giant component in Bernoulli bond percolation. As $\kappa$ decreases, the agreement in mean size is maintained, but the size distribution becomes broader, as evidenced by the increase in variance. This illustrates the breakdown of isomorphism and shows that the equality of the major epidemic size only holds in terms of expectation. Microscopically, under stronger correlation between outgoing transmissions, transmission from an infected node to multiple susceptible neighbors is more likely to be either all success or all failure. Due to the effect of such joint probability, the giant out-component size exhibits larger fluctuations.

\section{Discussions and conclusions}
The mapping between non-recurrent epidemics and percolation processes provides an analytically tractable instrument for studying the spread of infectious diseases in networks. In particular, the mapping to Bernoulli bond percolation is routinely used to simplify analyses, either unquestioningly or with an ad hoc justification, despite being shown to be imperfect. In this work, we asked precisely when this assumption is warranted. We distinguished three notions of equivalence between a general non-recurrent epidemic model and Bernoulli bond percolation---isomorphism (type 1), equality of node infection probabilities (type 2), and equality of the expected size of a major epidemic in large configuration model networks (type 3)---and identified the conditions for each.

Isomorphism requires symmetry, full independence, and equiprobability of transmission across all ordered pairs of adjacent nodes. Contrary to the common belief that isomorphism is possible only when the infection-to-recovery intervals are constant and equal, we showed that the conditions for isomorphism can be relaxed when infection-to-contact intervals are not i.i.d.  
Type-2 equivalence on trees and type-3 equivalence demand less: independence of outgoing transmissions is not needed, so Bernoulli bond percolation may remain accurate even when it is not isomorphic to the epidemic model. However, as soon as the contact network contains cycles---whether short cycles within a neighborhood or long cycles in an otherwise locally treelike network---the conditions for type-2 equivalence fall back to those for isomorphism. We note that the classical results for i.i.d.\ infection-to-contact intervals~\cite{newman2002spread, kenah2007second} are subsumed in type-3 equivalence as the special case where correlations between outgoing transmissions are induced by non-degenerate infection-to-recovery intervals.

Our numerical experiments quantify the discrepancies that arise when these conditions are violated. Correlations among outgoing, incoming, and transitive transmissions, as well as the presence of short cycles (triangles), each perturb node infection probabilities, but in different ways. The disparity due to correlated outgoing transmissions mainly stems from the reachability between the seeds and the giant strongly connected component; therefore, it vanishes once the seeds are conditioned to lie in the giant. Correlations among incoming and transitive transmissions, by contrast, render all three types of equivalence infeasible. Short cycles amplify the discrepancies already created by correlations in the epidemic dynamics. 
The effects of correlations are not linear with respect to correlation strength. We consistently observe threshold behavior, in which the deviation from Bernoulli bond percolation remains negligible while correlations are weak, and grows monotonically only once they exceed a certain magnitude. One caveat, though, is that the threshold value likely depends on the network structure, the marginal probability of transmission along a single edge, and the number of seed nodes. We will leave the systematic investigation to future work.

Type-2 and type-3 equivalences ensure that Bernoulli bond percolation accurately predicts the marginal probability that each node becomes infected during an outbreak or in a major epidemic. However, the joint probabilities are only preserved under isomorphism. This is clearly illustrated by the distributions of the largest out-component size in the correlated outgoing transmission model: The prediction of the expected size remains accurate regardless of the correlation strength, while its variance increases with stronger correlation.

Another important takeaway is that transitive correlations can increase node infection probabilities and, consequently, the expected outbreak size. Previous literature has held that the expected size and probability of a major epidemic are upper-bounded by their values under isomorphism and are not increased by the presence of correlations~\cite{kuulasmaa1982spatial, kenah2007second, miller2007epidemic, miller2008bounding, wilkinson2017relationships}. However, these results only concern node-level variabilities, such as the infection-to-recovery interval or independently distributed infectivity and susceptibility, that induce correlations among outgoing or incoming transmissions in our framework. By accounting for a broader class of correlation structures, we show that the effect of correlations is not unidirectional; transitive correlations can in fact increase the expected size of an outbreak or a major epidemic.

The modeling considerations we adopt here are not merely of theoretical interest. In particular, we deliberately avoid the typical i.i.d.\ postulate because infection-to-contact and infection-to-recovery intervals can be (and often are) correlated in real-world transmission processes due to the following mechanisms.

First, the latent period---the delay between contracting the infection and becoming capable of transmitting it---bounds both intervals from below. An individual who takes longer to become infectious will, all else equal, both make their first transmissible contact later and recover later, inducing a positive correlation between the two intervals. 

Second, the correlation may be induced by the within-host trajectory of the pathogen. Infectiousness tracks pathogen load, which typically rises after infection, peaks, and then declines as the immune response clears it. Therefore, the same trajectory governs both when transmissible contacts occur and how long the individual remains infectious. 
When an infected individual has a higher peak pathogen load, the contacts they make are likely to become transmissible sooner after the infection. Such individuals also tend to experience more prolonged infection due to slower clearance, which leads to a negative correlation between the infection-to-contact and infection-to-recovery intervals. Vaccination can have a similar effect because it attenuates infectiousness, which on average delays transmission while shortening the duration of infection.

Third, host behavior may introduce another type of coupling. When the appearance of symptoms prompts an infected individual to avoid the contacts through which the disease spreads, it effectively terminates the infectious period. From a modeling perspective, the onset of symptoms in such a case can be viewed as an effective recovery. Since the onset of infectiousness and the onset of symptoms are both outcomes of the same within-host progression, individuals who become infectious earlier may also withdraw earlier, leading again to a positive association.
Through these mechanisms, the two intervals may be entangled rather than distributed independently.

In Sec.~\ref{sec:nonequiv}, we further considered correlations between transmissions on different ordered pairs of nodes. These correlations could have multiple origins in real-world settings. 
For example, heterogeneity in individual infectivity, driven by differences in viral shedding, respiratory activity, or contact intensity, means that a highly infectious host is more likely to transmit to all of its neighbors, producing correlations among the outgoing transmissions of each node. 
Symmetrically, heterogeneity in susceptibility, associated with age, immunocompetence, and vaccination status, makes some individuals more readily infected upon contact. This leads to correlations between incoming transmissions. 
Correlations along transitive edges arise when the factors governing successive transmissions are themselves related. Assortative mixing, or homophily, provides one possible explanation. If individuals are more likely to be in contact with others of similar age, behavior, or immune status, then the attributes that influence transmissibility across consecutive edges are correlated. Shared context may also play a role. Transmissions in the same household, workplace, or environmental setting are subject to common conditions such as ventilation, crowding, and seasonality, which modulate the transmissibility of all contacts taking place there. Finally, pathogen evolution can introduce correlations: The variant passed from $i$ to $j$ is the same variant that $j$ may pass to $k$, and if this variant happens to be more or less transmissible than other variants, the transmission along transitive edges will be correlated. (Strictly speaking, however, invoking pathogen evolution is at odds with the assumption of time homogeneity that underlies our model.)

Our findings suggest that microscopic, individual-level correlations in transmission processes can have macroscopic consequences on epidemic dynamics. However, empirical evidence on these correlations is still limited. Knowledge of the correlation structure would provide valuable guidance on when mapping to percolation can be trusted, and when it is likely to fail.

\begin{acknowledgments}
    This work was supported by the Strategic Research Council (SRC) established within the Research Council of Finland (Decision Numbers: 364386 and 364371).
\end{acknowledgments}

\appendix
\section{Proof of the degeneracy of infection-to-recovery intervals}
\label{app:proof}

Here we prove our claim in Sec.~\ref{sec:isomorphism}: if the infection-to-contact intervals $\{T_{i \to j}\}$ are mutually independent and independent of all infection-to-recovery intervals $\{U_i\}$, and if the cumulative distribution function $F_{T_{i \to j}}$ is strictly increasing over the support of $U_i$, then the isomorphism conditions can be satisfied with a nontrivial retention probability $q \in (0,1)$ only if each $U_i$ is degenerate, i.e., $f_{U_i}(u) = \delta(u - u_i^*)$.

Consider a node $i$ with two neighbors $j_1, j_2$. We simplify the notation by letting $T_1 = T_{i \to j_1}$, $T_2 = T_{i \to j_2}$, and $U = U_i$. Given the independence of $T_m \, (m=1,2)$ and $U$, the probability of transmission from $i$ to $j_m$ is expressed as
\begin{equation}
    \begin{aligned}
    P(\xi_{i \to j_m} = 1) &= P(T_m < U) \\
    &= \int_0^\infty f_U(u) F_{T_m}(u) \,du \\
    &= \langle F_{T_m}(U) \rangle,
    \end{aligned}
\end{equation}
where $\langle \cdot \rangle$ denotes the expectation with respect to $U$. 
Due to the independence between $T_1$ and $T_2$, the joint probability of transmission to both neighbors is given by
\begin{equation}
    \begin{aligned}
    P(\xi_{i \to j_1} = 1, \xi_{i \to j_2} = 1) &= P(T_1 < U, T_2 < U) \\
    &= \int_0^\infty f_U(u) F_{T_1}(u) F_{T_2}(u) \,du \\
    &= \langle F_{T_1}(U) F_{T_2}(U) \rangle.
    \end{aligned}
\end{equation}
The independence condition for isomorphism requires the joint probability to factorize into the product of the marginals, leading to
\begin{equation}
    \langle F_{T_1}(U) F_{T_2}(U) \rangle = \langle F_{T_1}(U) \rangle \langle F_{T_2}(U) \rangle.
\end{equation}
In other words, the covariance between $F_{T_1}(U)$ and $F_{T_2}(U)$ must be zero. Let $U'$ be an independent copy of $U$. Then, the vanishing covariance implies
\begin{equation}
    \label{eq:covariance}
    \langle [F_{T_1}(U) - F_{T_1}(U')][F_{T_2}(U) - F_{T_2}(U')] \rangle = 0.
\end{equation}
Since $F_{T_1}$ and $F_{T_2}$ are non-decreasing functions, the product $[F_{T_1}(U) - F_{T_1}(U')][F_{T_2}(U) - F_{T_2}(U')]$ is non-negative for any realizations of $U$ and $U'$. Moreover, since we assume that $F_{T_1}$ and $F_{T_2}$ are strictly increasing over the support of $U$, this product is strictly positive whenever $U \neq U'$. Therefore, the lhs of Eq.~\eqref{eq:covariance} can be zero only if $U$ is degenerate, i.e., $P(U = u^*) = 1$, where $u^*$ is the constant that satisfies $F_{T_1}(u^*) = F_{T_2}(u^*) = q$. It is easy to see that this argument can be extended to any number of neighbors. Thus, the degeneracy of $U_i$ is a necessary condition for the independence of outgoing transmissions, and hence for isomorphism.


%

\end{document}